\documentclass[fleqn,12pt,twoside]{article}
\usepackage{espcrc1}

\usepackage{graphicx}
\usepackage[figuresright]{rotating}

\newcommand{\AmS}{{\protect\the\textfont2
  A\kern-.1667em\lower.5ex\hbox{M}\kern-.125emS}}

\newcommand{\Lower}[1]{\smash{\lower 1.5ex \hbox{#1}}}
\newcommand{\LN}{$\Lambda N$}
\newcommand{\LL}{$\Lambda\Lambda$}
\newcommand{\SN}{$\Sigma N$}

\newcommand{\LNSN}{$\Lambda N-\Sigma N$}
\newcommand{\LLNX}{$\Lambda\Lambda-N\Xi$}

\newcommand{\LS}{$\Lambda\Sigma$}

\newcommand{\NX}{$N\Xi$}
\newcommand{\SigSig}{$\Sigma\Sigma$}
\newcommand{\BL}{$B_\Lambda$}
\newcommand{\BLL}{$B_{\Lambda\Lambda}$}

\newcommand{\PS}{$P_\Sigma$}
\newcommand{\HII}{$^2$H}
\newcommand{\HIII}{$^3$H}
\newcommand{\HeIII}{$^3$He}
\newcommand{\HeIV}{$^4$He}
\newcommand{\HIIIL}{$_\Lambda^3$H}

\newcommand{\HIVL}{$_\Lambda^4$H}

\newcommand{\HeIVL}{$_\Lambda^4$He}

\newcommand{\HIVLs}{$_\Lambda^4$H$^\ast$}
\newcommand{\HeIVLs}{$_\Lambda^4$He$^\ast$}
\newcommand{\HeVL}{$_\Lambda^5$He}
\newcommand{\HIVLL}{$_{\Lambda\Lambda}^{\ \ 4}$H}

\newcommand{\HVLL}{$_{\Lambda\Lambda}^{\ \ 5}$H}
\newcommand{\HeVLL}{$_{\Lambda\Lambda}^{\ \ 5}$He}
\newcommand{\HeVILL}{$_{\Lambda\Lambda}^{\ \ 6}$He}

\title{Full-coupled channel approach to $S=-2$ 
$s$-shell hypernuclear systems 
}

\author{H. Nemura\address[KEK]{Institute of Particle and Nuclear 
Studies, KEK, \\
 Tsukuba 305-0801, Japan}, %
 S. Shinmura\address[Gifu]{Department of Information Science, 
 Gifu University, \\
 Gifu 501-1193, Japan},
 Y. Akaishi\addressmark[KEK] 
 and 
 Khin Swe Myint\address[Mandalay]{Department of Physics, 
 Mandalay University, \\
 Mandalay, Union of Myanmar}
}
       
\begin{document}

\maketitle

\begin{abstract}
 We describe full-coupled-channel {\it ab initio} calculations
 among the octet baryons for $S=-2$ $s$-shell hypernuclei, \HIVLL, 
 \HVLL\ %
 and \HeVILL. 
 The wave function includes \LL, \LS, \NX\ and \SigSig\ channels. 
 Minnesota $NN$, D2$^\prime$ $YN$ and Nijmegen model D simulated $YY$ 
 interactions are used. 
 This is the first attempt to explore the few-body problem of the 
 full-coupled channel scheme for $A=4-6$, $S=-2$ multistrangeness 
 hypernuclear systems. 
 Bound state solutions of the \LL\ hypernuclei, 
 \HIVLL, \HVLL\ and \HeVILL, are obtained. 
\end{abstract}

\section{INTRODUCTION --- Why full-coupled channel? ---}

Both experimental and theoretical searches for \HIVLL\ are 
of utmost interest 
in the field of 
hypernuclei\cite{Ahn,Kumagai,FG,Nem03,Pile,Kahana}. %
On the theoretical side, one of the conclusions in our
recent publication\cite{Nem03} is that, ``{\it
$^3S_1$ \LN\ interaction has to be determined very carefully, since 
\BLL\ is sensitive to the $^3S_1$ channel of the \LN\ interaction.
}''
Our standpoint can be explained by examining the relative importance of
$^3S_1$ and $^1S_0$ \LN\ and of $^1S_0$ \LL\ interactions in a simple
$core\ nucleus+2\Lambda$ model: 
\begin{equation}
\left\langle \sum_{i=1}^{N}\sum_{j=1}^{Y}v_{N_i \Lambda_j}\right\rangle 
= {\cal N}_t\ \bar{v}_{t,N\Lambda} + {\cal N}_s\ \bar{v}_{s,N\Lambda}, \qquad 
\left\langle \sum_{i<j}^{Y}v_{\Lambda_i \Lambda_j} \right\rangle
= n_s\ \bar{v}_{s,\Lambda\Lambda}, 
\end{equation}
where $N$ ($Y$) is the number of nucleons (hyperons) and thus $A=N+Y$. 
${\cal N}_t$, ${\cal N}_s$ and $n_s$ are the number of $^3S_1$ \LN, 
$^1S_0$ \LN\ and  
$^1S_0$ \LL\ pairs, respectively. 
Each $\bar{v}$ is the average of the potential matrix element in the 
appropriate channel. 
Table~\ref{spisofactor} lists the number of pairs, ${\cal N}_t$, 
${\cal N}_s$ and $n_s$ (right side). 
The table also lists the numbers for $S=-1$ hypernuclei (left side). 
In the $S=-2$ systems, the number $n_s$ is always $1$, 
and the ratio between ${\cal N}_t$ and ${\cal N}_s$ is always 
${\cal N}_t:{\cal N}_s=3:1$.  
This means that the \BLL\ binding energy strongly depends on the 
$^3S_1$ \LN\ interaction than the $^1S_0$ \LN\ or the 
$^1S_0$ \LL\ interaction. 
Particularly, in searching for \HIVLL, a check of the \LN\ potential 
concerning the observed binding energy of only the subsystem, \HIIIL, 
is insufficient, since the \BL(\HIIIL) strongly depends on the $^1S_0$ 
\LN\ interaction than the $^3S_1$ \LN\ interaction. 
The algebraic structure of the \LN\ pairs 
for $S=-2$ systems is very similar to the 
structure of \HeVL\ in $S=-1$ systems. 
This implies that the \LN\ interaction utilized in the study of $S=-2$ 
hypernuclei should reproduce the \BL(\HeVL) as well as the \BL's of 
$A=3,4$ $S=-1$ hypernuclei. 
However, a single channel \LN\ potential, 
which reproduces the \BL\ values of $A=3, 4$ $S=-1$ hypernuclei 
as well as the $\Lambda p$ total cross section, 
cannot reproduce the experimental \BL(\HeVL) value. 
This is known as a \HeVL\ anomaly, 
which was the long standing problem since the publication in 1972 by 
Dalitz {\it et al.}\cite{DHT}. 
According to the recent study by Akaishi {\it et al.}\cite{Akaishi}, 
$\Sigma$ degrees of freedom have to be explicitly taken into account 
so as to resolve the \HeVL\ anomaly.

Considering the fact that the \LL\ system couples to \NX\ and 
$\Sigma\Sigma$ states, as well as the \LN\ couples to the \SN, 
a theoretical search for \HIVLL\ should be made in a 
full-coupled channel formulation with a set of interactions 
among the octet baryons. 
Thus the purpose of this study is to describe a systematic study 
for 
$s$-shell \LL\ hypernuclei 
in a framework of fully coupled channel formulation.

\begin{table}[t]
\caption{Numbers of $^3S_1$ and $^1S_0$ \LN\ and $^1S_0$ \LL\ 
 pairs for $S=-1$ (left side) or $S=-2$ (right side) hypernucleus in a 
 ``$core\ nucleus + \Lambda$'' or a ``$core\ nucleus + 2\Lambda$'' model. 
}
\label{spisofactor}
\newcommand{\m}{\hphantom{$-$}}
\newcommand{\cc}[1]{\multicolumn{1}{c}{#1}}
\renewcommand{\tabcolsep}{1.8pc} 
\renewcommand{\arraystretch}{1.2} 
\begin{minipage}[t]{70mm}
\begin{tabular}{@{}lcc}
\hline
$S=-1$ & ${\cal N}_t$ & ${\cal N}_s$ \\
\hline
 \HIIIL & $1\over 2$ & $3\over 2$ \\
 \HIVL, \HeIVL & $3\over 2$ & $3\over 2$ \\
 \HIVLs, \HeIVLs & $5\over 2$ & $1\over 2$ \\
 \HeVL & $3$ & $1$ \\
\hline
\end{tabular}\\[2pt]
\end{minipage}
\begin{minipage}[t]{80mm}
\begin{tabular}{@{}lccc}
\hline
$S=-2$ & ${\cal N}_t$ & ${\cal N}_s$ & $n_s$\\
\hline
 \HIVLL & $3$ & $1$ & $1$ \\
 \HVLL, \HeVLL & $9\over 2$ & $3\over 2$ & $1$ \\
 \HeVILL & $6$ & $2$ & $1$ \\
\hline
\end{tabular}\\[2pt]
\end{minipage}
\end{table}

\section{INTERACTIONS AND METHOD}

The wave function of a system with strangeness $S=-2$, 
comprising $A$ octet baryons, has four isospin-basis components. 
For example, \HeVILL\ has four components as $ppnn\Lambda\Lambda$, 
$NNNNN\Xi$, $NNNN\Lambda\Sigma$ and $NNNN\Sigma\Sigma$. 
We abbreviate these components as \LL, \NX, \LS\ and $\Sigma\Sigma$, 
referring the last two baryons. 
The hamiltonian of the system is hence given by $4\times 4$ components 
as 
\begin{equation}
 H=\left(\begin{array}{cccc}
    H_{\Lambda\Lambda} & V_{N\Xi-\Lambda\Lambda} & 
     V_{\Lambda\Sigma-\Lambda\Lambda} & V_{\Sigma\Sigma-\Lambda\Lambda} \\
	  V_{\Lambda\Lambda-N\Xi} & H_{N\Xi} & 
	   V_{\Lambda\Sigma-N\Xi} & V_{\Sigma\Sigma-N\Xi} \\
	  V_{\Lambda\Lambda-\Lambda\Sigma} & V_{N\Xi-\Lambda\Sigma} & 
	   H_{\Lambda\Sigma} & V_{\Sigma\Sigma-\Lambda\Sigma} \\
	  V_{\Lambda\Lambda-\Sigma\Sigma} & V_{N\Xi-\Sigma\Sigma} &
	   V_{\Lambda\Sigma-\Sigma\Sigma} & H_{\Sigma\Sigma} \\
\end{array}
\right),
\end{equation}
where $H_{B_1B_2}$ operates on the $B_1B_2$ component, 
and $V_{B_1B_2-B_1^\prime B_2^\prime}$ is the sum of 
all possible two-body transition potential connecting 
$B_1B_2$ and $B_1^\prime B_2^\prime$ components. 

In the present calculation, we use Minnesota potential
for the $NN$ interaction, D2$^\prime$ for the $YN$ and Nijmegen model D 
simulated (ND(S)) for the $YY$ interaction. 
The Minnesota potential reproduces reasonably well both the binding 
energies and sizes of few-nucleon systems, 
such as \HII, \HIII, \HeIII\ and \HeIV\cite{PreciseVS}. 
The D2$^\prime$ potential is a modified potential from the original D2 
potential\cite{Akaishi}.
The strength of the long-range part 
($V_b$ in Table I of Ref.~\cite{Akaishi}) of 
the D2$^\prime$ potential in the \LN-\LN\ $^3S_1$ 
channel is reduced by multiplying a factor $0.954$, in order to reproduce 
the experimental \BL(\HeVL) value. 
We take the hard core radius as $r_c=0.56(0.45)$ fm for the 
ND 
$YY$ interaction
in the $^1S_0(^3S_1)$ channel, 
which is the same as the hard core radius of the $YN$ sector. 
The ND(S) potential is given by 
\begin{equation}
v(r) = v_{S}\exp\{-(r/\beta_S)^2\} + v_{L}\exp\{-(r/\beta_L)^2\},
\end{equation}
with $\beta_S=0.5$ fm and $\beta_L=1.2$ fm. 
The strength parameters are shown in Table~\ref{NDSPOT}. 

The calculations were made by using 
stochastic variational method\cite{PreciseVS}. 
The reader is referred to Ref.~\cite{abinitio} for the 
details of the method.

\begin{table}[t]
\caption{
 Parameters of the ND(S) $YY$ potential in the even states, 
 given in units of MeV. 
}
 \label{NDSPOT}
\newcommand{\m}{\hphantom{$-$}}
\newcommand{\cc}[1]{\multicolumn{1}{c}{#1}}
\renewcommand{\arraystretch}{1.2} 
\begin{tabular}{@{}ccrrcrr}
\hline
 & & \multicolumn{2}{c}{$^1E$} & & \multicolumn{2}{c}{$^3E$} \\
\cline{3-4} \cline{6-7}
 & & $v_S$\quad~ & $v_L$\quad~ & & $v_S$\quad~ & $v_L$\quad~ \\
\hline
 \LL$-$\LL & $I=0$ & $1464.48$ & $-95.00$ & & ~    & ~    \\
 \LL$-$\NX & $I=0$ & $200.86$ & $15.57$ & & ~    & ~ \\
 \NX$-$\NX & $I=0$ & $1205.68$ & $-89.19$ & & $968.12$ & $-70.28$ \\
 ~         & $I=1$ & $1783.17$ & $-66.54$ & & $862.75$ & $-72.23$ \\
 \LL$-$\SigSig & $I=0$ & $648.96$ & $-42.81$ & & ~ & ~ \\
 \NX$-$\LS     & $I=1$ & $103.84$ & $62.98$ & & $-311.72$ & $-42.42$ \\
 \NX$-$\SigSig & $I=0$ & $-301.11$ & $87.11$ & & ~ & ~ \\
 ~             & $I=1$ & ~ & ~ & & $-14.18$ & $-8.26$ \\
 \LS$-$\LS     & $I=1$ & $1245.45$ & $-90.67$ & & $909.99$ & $-85.40$ \\
 \LS$-$\SigSig & $I=1$ & ~ & ~ & & $-129.02$ & $-41.88$ \\
 \SigSig$-$\SigSig & $I=0$ & $350.73$ & $-15.32$ & & ~ & ~ \\
 ~                 & $I=1$ & ~ & ~ & & $738.86$ & $-74.71$ \\
 ~                 & $I=2$ & $693.54$ & $-114.26$ & & ~ & ~ \\
\hline
\end{tabular}\\[2pt]
\end{table}

\section{RESULTS AND DISCUSSIONS}

\begin{table}[b]
\caption{$\Lambda$ and \LL\ separation energies, given in units of MeV, 
 of $A=3-6$, $S=-1$ and $-2$ $s$-shell hypernuclei. 
}
 \label{BEA3456}
\newcommand{\m}{\hphantom{$-$}}
\newcommand{\cc}[1]{\multicolumn{1}{c}{#1}}
\renewcommand{\arraystretch}{1.2} 
\begin{tabular}{@{}lccccccc}
\hline
 & \BL(\HIIIL) & \BL(\HIVL) & \BL(\HIVLs) & \BL(\HeVL) 
 & \BLL(\HIVLL) & \BLL(\HVLL) & \BLL(\HeVILL) \\
\hline
 Calc & $0.056$ & $2.23$ & $0.91$ & $3.18$ 
 & $0.107$ & $4.04$ & $7.93$ \\
 Exp  & $0.13(5)$ & $2.04(4)$ & $1.00(4)$ & $3.12(2)$ 
 & & & $7.3(3)$ \\
\hline
\end{tabular}\\[2pt]
\end{table}

\begin{table}[t]
\caption{Probabilities, given in percentage, of %
 \NX, \LS\ and \SigSig\ components for \LL\ hypernuclei, 
 $_{\Lambda\Lambda}^{\ A}$Z. 
 The probabilities of the $\Sigma$ component for the subsystems, 
 $_{\ \ \ \Lambda}^{A-1}$Z, %
 are also given in parentheses. 
}
 \label{ProbA3456}
\newcommand{\m}{\hphantom{$-$}}
\newcommand{\cc}[1]{\multicolumn{1}{c}{#1}}
\renewcommand{\tabcolsep}{2pc} 
\renewcommand{\arraystretch}{1.2} 
\begin{tabular}{@{}llll}
\hline
 & \HIVLL\ (\HIIIL) & \HVLL\ (\HIVL, \HIVLs)& \HeVILL\ (\HeVL) \\
\hline
 $P_{N\Xi}$ %
 & $0.12$ & $4.34$ & $0.27$ \\
 $P_{\Lambda\Sigma}$ (\PS) %
 & $0.35$ ($0.16$) & $2.52$ ($2.17$, $0.36$)
 & $1.18$ ($0.55$) \\ 
 $P_{\Sigma\Sigma}$ %
 & $0.01$ & $0.05$ & $0.04$ \\
\hline
\end{tabular}\\[2pt]
\end{table}

Table~\ref{BEA3456} lists 
the \BL\ and \BLL\ values for $S=-1$ and 
$-2$ hypernuclei. 
The D2$^\prime$ $YN$ potential well reproduces all the \BL\ values for 
$A=3-5$, $S=-1$ hypernuclei. 
The D2$^\prime$ includes explicit \LNSN\ coupling, which is %
significantly important %
to resolve the~\HeVL~anomaly. 

Using the ND(S) $YY$ potential, we have obtained the bound state 
solutions of \HIVLL, \HVLL\ and \HeVILL. 
The obtained \BLL(\HeVILL) is slightly larger than the experimental 
value; 
$\Delta B_{\Lambda\Lambda}^{\mbox{(calc)}}(\mbox{\HeVILL})=1.58$ MeV, 
while $\Delta B_{\Lambda\Lambda}^{\mbox{(exp)}}(\mbox{\HeVILL})=1.01\pm
0.20^{+0.18}_{-0.11}$ MeV\cite{Nagara}. 
The calculated $\mbox{\BLL(\HIVLL)}=0.107$ MeV is very close to but 
lower than the obtained $\mbox{\HIIIL}+\Lambda$ threshold, 
$\mbox{\BL(\HIIIL)}=0.056$ MeV.

Table~\ref{ProbA3456} lists the probabilities of $\Sigma$ for $S=-1$ and 
of \NX, \LS\ and \SigSig\ components for $S=-2$ hypernuclei. 
In the present calculation, the \LL\ component is the main part of the 
wave function of the $S=-2$ hypernuclei. 
No unrealistic bound states were found, since the hard core model hardly 
incorporates unrealistic strong attractive force in the short range 
region, in contrast to the soft core model such as NSC97e or 
NSC97f\cite{Yamada}. 
This is one of the reasons why we used the hard core model 
for the $YY$ interaction for the first attempt to the 
full-coupled channel calculation. 
$P_{\Sigma\Sigma}$'s are very small %
for both systems, 
due to large mass difference between \LL\ and \SigSig\ channels
($m_{\Sigma\Sigma}-m_{\Lambda\Lambda}\cong 155$ MeV). 

The $P_{N\Xi}(\mbox{\HVLL})$ is a surprisingly large amount, 
in spite of the fact that %
the strength of the \LLNX\ coupling of ND is not strong. 
This implies that the coupling between $t+\Lambda+\Lambda$ channel and 
$\alpha+\Xi^-$ channel ($\alpha$ formation effect\cite{KSMSA}) is 
significant.

We should note that the present results depend on the 
choice of the $YY$ potential. 
If we use other $YY$ potential model, 
the quantitative results must change. 
More comprehensive study using various kinds of $YY$ interaction models 
will be reported in the near future.

\end{document}